\documentclass[aps,prb,reprint,twocolumn,superscriptaddress,showpacs,floatfix,longbibliography]{revtex4-1}

\usepackage{amsmath}
\usepackage{color}
\usepackage{mathrsfs}
\usepackage{dcolumn}
\usepackage{bm}
\usepackage{multirow}
\usepackage{graphicx}
\usepackage{rotating}
\usepackage{layout}
\usepackage[version=3]{mhchem}
\usepackage{braket}
\usepackage{longtable}
\usepackage{setspace}

\allowdisplaybreaks
\usepackage{xfrac}

\usepackage[colorlinks,citecolor=blue,urlcolor=blue,bookmarks=false,hypertexnames=true]{hyperref}

\begin{document}

\title{Accurate prediction of K-edge excitation energies using
state-specific self-consistent perturbation theory}
\date{\today}
\author{Lan Nguyen Tran}
\email{tnlan@hcmus.edu.vn}
\affiliation{University of Science, Vietnam National University, Ho Chi Minh City 700000, Vietnam}
\affiliation{Vietnam National University, Ho Chi Minh City 700000, Vietnam}
\date{\today}

\begin{abstract}
We present the application of the recently developed one-body M{\o}ller--Plesset perturbation theory (OBMP2) to the prediction of K-edge excited states. OBMP2 is a self-consistent perturbation theory in which a canonical transformation followed by a cumulant approximation yields an effective one-body Hamiltonian. This resulting operator augments the standard Fock operator with a one-body correlation potential containing double-excitation MP2 amplitudes, allowing molecular orbitals and orbital energies to be optimized in the presence of correlation. This self-consistent framework mitigates convergence and accuracy issues often encountered in standard non-iterative MP2 for open-shell systems and bond-stretching regimes. In this work, we evaluate the performance of an OBMP2-based approach for the calculation of K-edge excitations. Utilizing benchmark test sets of both closed-shell and open-shell molecules, we demonstrate that our method outperforms established standard techniques, including $\Delta$DFT, EOM-CCSD, and USTEOM-CCSD. Our findings establish the OBMP2-based $\Delta$SCF protocol as a robust and accurate new computational method for the treatment of K-edge excited states. 
\end{abstract}

\maketitle

\section{Introduction}
The accurate characterization of core-level excitations is a cornerstone of modern X-ray spectroscopy, offering unique element-specific insights into the local geometric and electronic structure of molecules and materials. K-edge absorption spectroscopy, involving the excitation of a $1s$ core electron to unoccupied valence or Rydberg orbitals, has become an indispensable tool across a wide spectrum of scientific areas \cite{Stohr1992}. For example, in materials science, X-ray absorption near-edge structure (XANES) provides important information about oxidation states and coordination environments in catalysts and battery materials \cite{bordiga2013}. Similarly, in biology and chemistry, K-edge spectroscopy is pivotal for elucidating the active sites of metalloproteins and tracking transient species in photochemical reactions \cite{Yano2006, Penfold2018}. As experimental facilities, such as X-ray free-electron lasers (XFELs) and high-brilliance synchrotrons, achieve unprecedented time and energy resolution, the demand for equally precise and efficient theoretical methods to interpret these complex spectra has attracted a great attention\cite{Norman2018}.

Theoretical modeling of K-edge excited states presents significant challenges compared to valence excited states due to the strong relaxation effects of core hole orbitals. Several standard computational approaches have been established to address these issues. Time-dependent density functional theory (TD-DFT) is perhaps the most widely used approach due to its favorable balance between accuracy and computational cost, scaling formally as $O(N^3)$ to $O(N^4)$ \cite{Besley2010}. However, standard TD-DFT often significantly underestimates core-excitation energies due to the lack of orbital relaxation and requires large empirical shifts to align with experiment \cite{Lopata2012}. 

Regarding linear-response wavefunction methods, equation-of-motion coupled cluster with singles and doubles (EOM-CCSD) is considered a gold standard for valence excited states \cite{Coriani2012}. However, as shown by previous authors, EOM-CCSD is unable to achieve high accuracy for K-edge excited states \cite{ranga2021core}. Also, the steep scaling of EOM-CCSD, $O(N^6)$, renders it computationally prohibitive for large biochemical systems or complex material interfaces. Other wavefunction perturbation approaches, such as second-order algebraic diagrammatic construction [ADC(2)] with the scaling of $O(N^5)$, offer a middle ground but still struggle with the state-specific orbital relaxation required for core-level excited states.

An alternative approach to excited-state treatment is the Delta-self-consistent field ($\Delta$SCF) method. Unlike linear response theories, $\Delta$SCF calculates excitation energies as the energy difference between two distinct self-consistent field solutions: the ground state and a non-Aufbau excited state where a core electron is promoted to a virtual orbital \cite{Besley2009}. The primary advantage of $\Delta$SCF is its explicit treatment of orbital relaxation; by optimizing the orbitals in the presence of the core hole, it captures the significant reorganization energy missing in linear response methods. Furthermore, it retains the low computational cost of ground-state Hartree-Fock and DFT. However, $\Delta$SCF has some major drawbacks. It is prone to the variational collapse, where the calculation falls back to the lower-energy ground state during optimization. Additionally, obtaining orthogonal excited states is theoretically challenging, and the method generally lacks a systematic way to treat multi-determinant character or spin contamination \cite{Hait2020}.

We have recently developed a novel self-consistent perturbation theory, termed one-body M{\o}ller–Plesset perturbation theory (OBMP2) \cite{OBMP2-JCP2013,OBMP2-JPCA2021, OBMP2-PCCP2022, OBMP2-JPCA2023}. In OBMP2, a canonical transformation \cite{CT-JCP2006,CT-JCP2007,CT-ACP2007,CT-JCP2009,CT-JCP2010,CT-IRPC2010} followed by a cumulant approximation \cite{cumulant-JCP1997,cumulant-PRA1998,cumulant-CPL1998,cumulant-JCP1999} yields an effective one-body Hamiltonian. The resulting operator comprises the standard Fock operator plus a one-body correlation potential that includes double-excitation MP2 amplitudes. Molecular orbitals and orbital energies are then optimized in the presence of correlation by diagonalizing a correlated Fock matrix. The self-consistency of OBMP2 mitigates issues associated with the non-iterative nature of standard MP2 for open-shell systems \cite{OBMP2-JPCA2021, OBMP2-PCCP2022} and for bond-stretching regimes \cite{OBMP2-JPCA2023}. In many cases, OBMP2 outperforms orbital optimized MP2 (OO-MP2) for open-shell systems \cite{OBMP2-PCCP2022}. We have further extended OBMP2 to a spin-opposite variant (O2BMP2) \cite{OBMP2-JPCA2024} and demonstrated that OBMP2 and O2BMP2 can treat intermolecular charge-transfer excited states with accuracy comparable to high-level coupled-cluster methods \cite{OBMP2-JCP2025}. In this work, we evaluate the efficacy of an OBMP2-based $\Delta$SCF approach for predicting K-edge excited states. Utilizing benchmark test sets of closed-shell and open-shell molecules 
we demonstrate that our method surpasses standard techniques, including $\Delta$DFT, EOM-CCSD, and USTEOM-CCSD. These results establish OBMP2-based $\Delta$SCF as a promising new computational tool for the accurate treatment of core-level excitations.   

\section{Review of OBMP2 theory}
A comprehensive derivation of the OBMP2 theoretical framework is provided in Refs.~\cite{OBMP2-JPCA2021,OBMP2-PCCP2022,OBMP2-JPCA2023,OBMP2-JPCA2024}; the method has been implemented within a locally modified version of the PySCF quantum chemistry package\cite{pyscf-2018}. The OBMP2 effective Hamiltonian is constructed via the canonical transformation formalism\cite{CT-JCP2006,CT-JCP2007,CT-ACP2007,CT-JCP2009,CT-JCP2010,CT-IRPC2010}:
\begin{align}
\hat{\bar{H}} = e^{\hat{A}^\dagger} \hat{H} e^{\hat{A}},
\label{Hamiltonian:ct}
\end{align}
where the molecular electronic Hamiltonian is expressed as
\begin{align}
  \hat{H} =  \sum_{pq}h^{p}_{q} \hat{a}_{p}^{q} + \tfrac{1}{2}\sum_{pqrs}g^{p r}_{q s}\hat{a}_{p r}^{q s}\label{eq:h1}.
\end{align}
In the above, the indices $\left\{p, q, r, \ldots \right\}$ denote general ($all$) spin orbitals. The one- and two-body second-quantized operators $\hat{a}_p^q$ and $\hat{a}_{pq}^{rs}$ are defined as $\hat{a}_p^q = \hat{a}^\dagger_p\hat{a}_q$ and $\hat{a}_{pq}^{rs} = \hat{a}^\dagger_p\hat{a}^\dagger_q\hat{a}_s\hat{a}_r$, respectively. The quantities $h_{pq}$ and $v_{pq}^{rs}$ represent the one- and two-electron integrals. Within the OBMP2 framework, the anti-Hermitian cluster operator $\hat{A}$ is restricted to double excitations:
\begin{align}
  \hat{A} = \hat{A}_\text{D} = \tfrac{1}{2} \sum_{ij}^{occ} \sum_{ab}^{vir} T_{ij}^{ab}(\hat{a}_{ab}^{ij} - \hat{a}_{ij}^{ab}) \,, \label{eq:op1}
\end{align}
where the amplitudes are determined at the second-order M{\o}ller--Plesset (MP2) level:
\begin{align}
  T_{i j}^{a b} =  \frac{g_{i j}^{a b} } { \epsilon_{i} + \epsilon_{j} - \epsilon_{a} - \epsilon_{b} } \,, \label{eq:amp}
\end{align}
with the indices $\left\{i, j, k, \ldots \right\}$ labeling occupied ($occ$) spin orbitals, $\left\{a, b, c, \ldots \right\}$ labeling virtual ($vir$) spin orbitals, and $\epsilon_{i}$ denoting the canonical orbital energy associated with spin-orbital $i$. Using Baker–Campbell–Hausdorff (BCH) expansion up to the second order for the effective Hamiltonian (Eq.~\ref{Hamiltonian:ct}) and employing some approximations, we arrive at the OBMP2 Hamiltonian as follows,

\begin{align}
  \hat{H}_\text{OBMP2} = \hat{H}_\text{HF} + \left[\hat{H},\hat{A}_\text{D}\right]_1 + \tfrac{1}{2}\left[\left[\hat{F},\hat{A}_\text{D}\right],\hat{A}_\text{D}\right]_1 \label{eq:h4}.
\end{align}
Where the subscript 1 on the commutators, $[\ldots]_1$, indicates that only one-body operators and scalar constants, obtained from the reduction of many-body operators via the cumulant decomposition\cite{cumulant-JCP1997,cumulant-PRA1998,cumulant-CPL1998,cumulant-JCP1999}, are retained. The operator $\hat{H}_\text{HF}$ denotes the standard Hartree--Fock Hamiltonian.  $\hat{v}_\text{OBMP2}$, composing of one- and second-order BCH terms, constitutes a correlation potential comprising exclusively one-body operators, whose explicit working expression reads
\begin{align}
\hat{v}_{\text{OBMP2}} = &\left[\hat{H},\hat{A}_\text{D}\right]_1 + \tfrac{1}{2}\left[\left[\hat{F},\hat{A}_\text{D}\right],\hat{A}_\text{D}\right]_1 \\
= &  \overline{T}_{i j}^{a b} \left[ f_{a}^{i} \,\hat{\Omega}\left( \hat{a}_{j}^{b} \right) 
  + g_{a b}^{i p} \,\hat{\Omega} \left( \hat{a}_{j}^{p} \right) - g^{a q}_{i j} \,\hat{\Omega} \left( \hat{a}^{b}_{q} \right) \right] \nonumber \\  &- 2 \overline{T}_{i j}^{a b}g^{i j}_{a b} 
   + \,f_{a}^{i}\overline{T}_{i j}^{a b}\overline{T}_{j k}^{b c} \,\hat{\Omega} \left(\hat{a}_{c}^{k} \right) \nonumber \\ 
     &+  f_{c}^{a}T_{i j}^{a b}\overline{T}_{i l}^{c b} \,\hat{\Omega} \left(\hat{a}^{l}_{j} \right) + f_{c}^{a}T_{i j}^{a b}\overline{T}_{k j}^{c b} \,\hat{\Omega} \left(\hat{a}^{k}_{i} \right) \nonumber \\ 
     &-  f^{k}_{i}T_{i j}^{a b}\overline{T}_{k l}^{a b} \,\hat{\Omega} \left(\hat{a}_{l}^{j} \right)
     -  f^{p}_{i}T_{i j}^{a b}\overline{T}_{k j}^{a b} \,\hat{\Omega} \left(\hat{a}^{p}_{k} \right) \nonumber \\ 
     & +  f^{k}_{i} T_{i j}^{a b}\overline{T}_{k j}^{a d} \,\hat{\Omega}\left(\hat{a}_{b}^{d} \right) +  f_{k}^{i}T_{i j}^{a b}\overline{T}_{k j}^{c b} \,\hat{\Omega} \left(\hat{a}_{a}^{c} \right) \nonumber \\ 
     &-  f_{c}^{a}T_{i j}^{a b}\overline{T}_{i j}^{c d} \,\hat{\Omega} \left(\hat{a}^{b}_{d} \right) \,
     - f_{p}^{a}T_{i j}^{a b}\overline{T}_{i j}^{c b} \,\hat{\Omega} \left(\hat{a}^{p}_{c} \right) \nonumber \\
     & - 2f_{a}^{c}{T}_{i j}^{a b}\overline{T}_{i j}^{c b} +  2f_{i}^{k}{T}_{i j}^{a b}\overline{T}_{k j}^{a b}. \label{eq:vobmp2} 
\end{align}
Here, $\overline{T}_{ij}^{ab} = {T}_{ij}^{ab} - {T}_{ji}^{ab}$ denotes the antisymmetrized amplitude, $\hat{\Omega} \left( \hat{a}^{p}_{q} \right) = \hat{a}^{p}_{q}  + \hat{a}^{q}_{p}$ is the symmetrization operator, and the Fock matrix elements are defined as
\begin{align}
    f_p^q = h_p^q + \sum_{i}^{occ}\left(g^{p i}_{q i} - g^{p i}_{i q} \right).
\end{align}
The OBMP2 Hamiltonian $\hat{H}_\text{OBMP2}$ (Eq.\ref{eq:h4}) can be reformulated in a manner analogous to the standard Hartree--Fock formalism:
\begin{align}
  \hat{H}_\text{OBMP2} = & \hat{\bar{F}} + \bar{C}, \label{eq:h5}
\end{align}
where $\bar{C}$ is a scalar constant aggregating all terms devoid of excitation operators. The quantity $\hat{\bar{F}}$ denotes the correlated Fock operator, $\hat{\bar{F}} =  \bar{f}^{p}_{q} \hat{a}_{p}^{q}$, with the correlated Fock matrix elements $\bar{f}^{p}_{q}$ expressed as
\begin{align}
\bar{f}^{p}_{q} &= f^{p}_{q} + v^{p}_{q}. \label{eq:corr-fock}
\end{align}
The correlation potential $v^{p}_{q}$ modifies the uncorrelated Hartree--Fock electronic structure. Consequently, the molecular orbital (MO) coefficients and orbital energies are obtained as the eigenvectors and eigenvalues, respectively, of the correlated Fock matrix $\bar{f}^{p}_{q}$.

To selectively converge upon a specific electronic state during the OBMP2 self-consistent procedure, we utilize the maximum overlap method (MOM) algorithm, originally developed for Hartree--Fock (HF) and density functional theory (DFT) calculations\cite{MOM-JPCA2008, MOM-JCTC2018}. The procedure begins with a MOM-HF calculation targeting the desired state, the solution of which subsequently serves as the initial reference for re-optimization within the OBMP2 and O2BMP2 frameworks. Additionally, the direct inversion of the iterative subspace (DIIS) technique\cite{diis1980} is employed to accelerate convergence. Unless otherwise noted, the aug-cc-pVTZ basis set \cite{Peterson2007augccpvxz} is used throughout this work.

\section{Results and discussion}

\begin{figure*}[t!]
  \includegraphics[width=12cm]{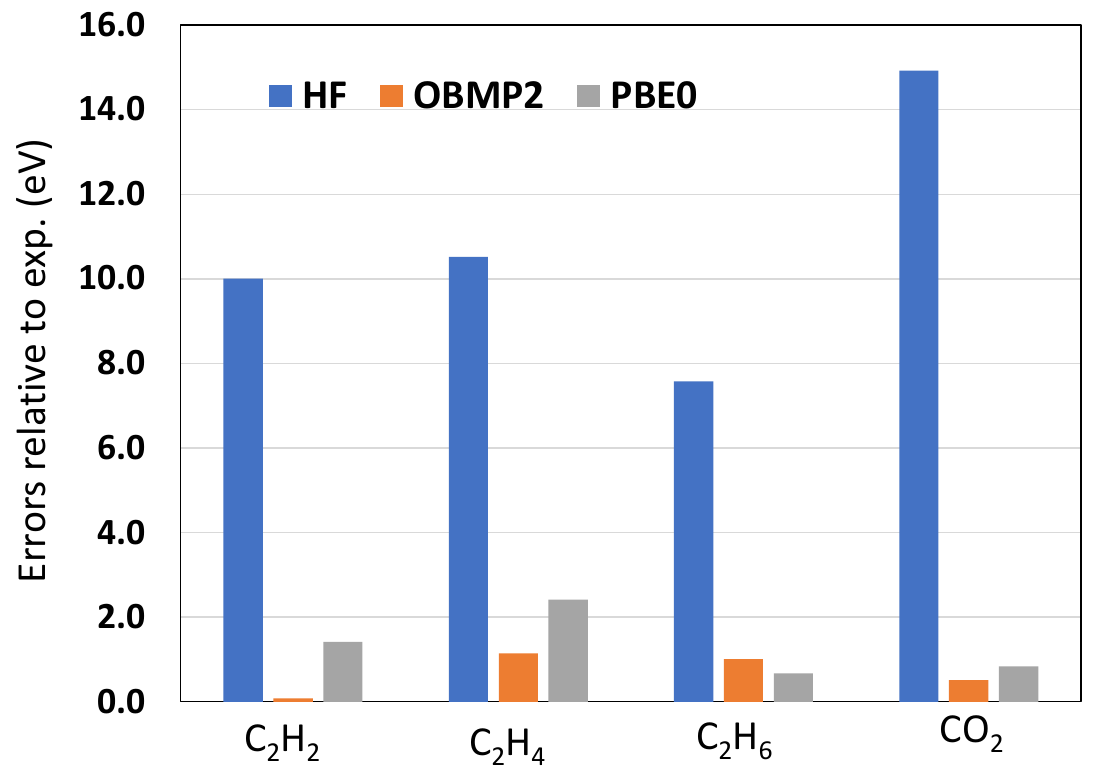}
  \caption{Errors relative to experiment of K-edge excitation energies from $\Delta$HF, $\Delta$OBMP2, and $\Delta$PBE0 for the C center in four molecules.}
  \label{fig:c_center}
\end{figure*}

Let us first evaluate the performance of OBMP2 on the K-edge excitation energies of closed-shell molecules. In Figure~\ref{fig:c_center}, we present the errors relative to experimental values for the K-edge excitation energies of the C center in C$_2$H$_2$, C$_2$H$_4$, C$_2$H$_6$, and CO$_2$, whose experimental values are adapted from Refs.~\cite{tronc1979carbon,hitchcock1977carbon}. It is evident that Hartree-Fock (HF) exhibits significant deficiencies, yielding substantial errors. In contrast, both OBMP2 and PBE0 produce much smaller errors, underscoring the importance of electron correlation for accurately predicting K-edge excitation energies. Overall, OBMP2 demonstrates smaller errors than PBE0, except in the case of C$_2$H$_6$, where the performance of OBMP2 and PBE0 is comparable.

\begin{figure*}[t!]
  \includegraphics[width=12cm]{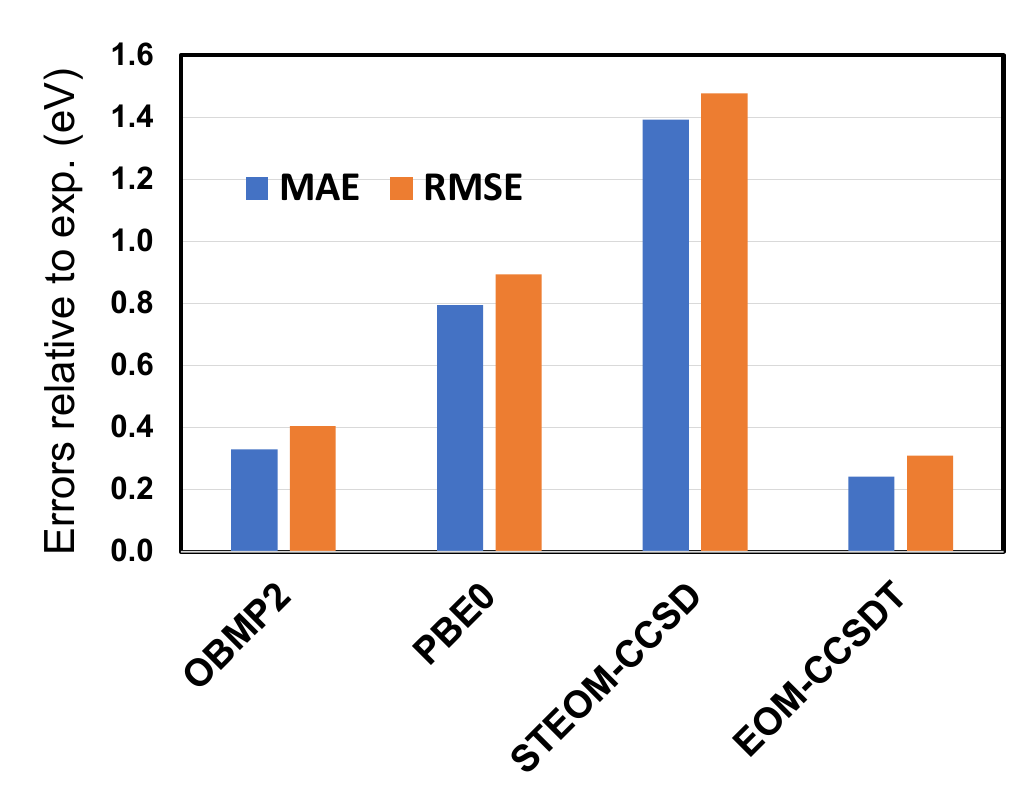}
  \caption{MAEs and RMSEs relative to experiment for different K-edge transition given in Table~\ref{tab:closed-shell}.}
  \label{fig:err_closed}
\end{figure*}

In Table~\ref{tab:closed-shell}, we report K-edge transitions to various valence orbitals for four molecules: CO, H$_2$O, HCN, and NH$_3$. In general, both OBMP2- and PBE0-based $\Delta$SCF methods predict K-edge excitation energies that align closely with experimental values. The statistical errors are summarized in Figure~\ref{fig:err_closed}. For comparison, we also report the mean absolute errors (MAEs) and root mean square errors (RMSEs) of CVS-EOM-CCSD and CVS-EOM-CCSDT, adapted from Refs.~\cite{matthews2020eom, jana1999development}. It is well established that significant orbital-relaxation effects must be accounted for when treating core-electron excitations; these effects cannot be adequately described by the linear excitation operator in EOM-CCSD. Consequently, EOM-CCSD exhibits substantial errors relative to experiment. The inclusion of triples in EOM-CC (EOM-CCSDT) markedly reduces these errors, yielding MAEs and RMSEs below 0.2 eV. Although $\Delta$DFT with the PBE0 functional yields K-edge excitation energies closer to experiment than EOM-CCSD, its errors remain relatively large, with an MAE and RMSE around 0.8 eV. Interestingly, $\Delta$OBMP2 attains an accuracy comparable to EOM-CCSDT, with an MAE and RMSE below 0.3 eV, several times smaller than $\Delta$PBE0. Overall, our results indicate that $\Delta$OBMP2 accurately predicts K-edge excitation energies for the small closed-shell molecules studied here, supporting the validity of our approach for K-edge excited states.

\begin{table}[t!]
  \normalsize
  \caption{K-Edge Excitation Energies (eV) of Closed-Shell Molecules Calculated from $\Delta$OBMP2 and $\Delta$PBE0. $^a$Experimental values adapted from Ref.~\cite{CO_Kedge} for CO, Ref.~\cite{H2O_NH3_Kedge} for H$_2$O and NH$_3$, and Ref.~\cite{HCN_Kedge} for HCN. \label{tab:closed-shell} \normalsize }
  \begin{tabular}{ccccccccccccccccccccc}
    \hline \hline		
    Molecules &Transition       &Exp$^a$	&OBMP2	&PBE0  \\
    \hline
C\textbf{O}	&K $\rightarrow \pi^*$	    &534.2	&534.4	&533.1 \\
	&K $\rightarrow$ 3s         &538.8	&539.2	&538.2 \\
	&K $\rightarrow$ $\pi^*$/3d	&539.8	&539.4	&539.2 \\
\textbf{C}O	&K $\rightarrow \pi^*$ 	    &287.4	&286.7	&286.1 \\
	&K $\rightarrow$ 3s 	    &292.5	&291.9	&292.4 \\
	&K $\rightarrow \pi^*$/3d	&293.4	&293.1	&293.5 \\
H$_2$\textbf{O}	&K $\rightarrow$ 3s	        &534.0	&534.2	&533.2 \\
	&K $\rightarrow$ 3p	        &535.9	&536.0	&534.8 \\
	&K $\rightarrow$ 4s	        &537.0	&537.6	&536.8 \\
HC\textbf{N}	&K $\rightarrow \pi^*$	    &399.7	&399.7	&398.8 \\
	&K $\rightarrow$ 3s  	    &402.5	&402.6	&403.1 \\
H\textbf{C}N	&K $\rightarrow \pi^*$	    &286.4	&286.4	&285.6 \\
	&K $\rightarrow$ 3s	        &289.1	&289.2	&288.7 \\
	&K $\rightarrow$ 4s	        &290.6	&289.7	&289.6 \\
\textbf{N}H$_3$	&K $\rightarrow$ 3s	        &400.8	&400.5	&400.0 \\
	&K $\rightarrow$ 3p	        &402.5	&402.2	&401.5 \\
	&K $\rightarrow$ 4s	        &403.0	&403.4	&401.5 \\
	&K $\rightarrow$ 5s	        &404.3	&404.1	&402.8 \\
\hline \hline
\end{tabular}
\end{table}

\begin{figure*}[t!]
  \includegraphics[width=12cm]{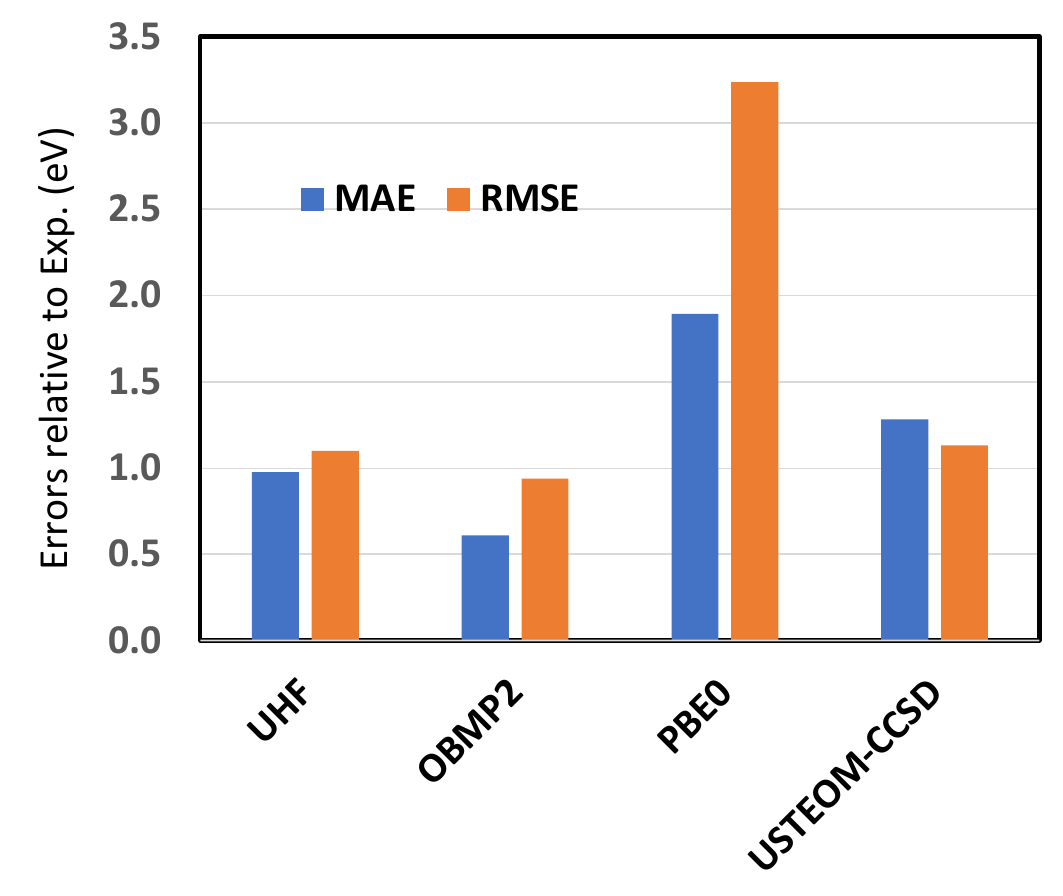}
  \caption{MAEs and RMSEs relative to experiment for different K-edge transition given in Table~\ref{tab:open-shell}.}
  \label{fig:err_open}
\end{figure*}

We now consider a series of small open-shell molecules and cations, consisting of NO, CO$^+$, OH, NH$_2^+$, and NH$^+$. Table~\ref{tab:open-shell} presents K-edge transition energies from various theoretical methods, UHF, OBMP2, PBE0, and USTEOM-CCSD. The statistical error metrics, specifically the mean absolute error (MAE) and root mean square error (RMSE), reveal distinct hierarchies in accuracy among the tested methods are shown in Figure~\ref{fig:err_open}. Overall, OBMP2 demonstrates the highest accuracy with a MAE of 0.6 eV and an RMSE of 0.9 eV, significantly outperforming the other approaches. In contrast, PBE0 exhibits the largest deviations, with a MAE of 1.9 eV and an RMSE of 3.2 eV, suggesting it struggles to capture core-level excitation dynamics in these specific systems.

Looking closely, we can see that $\Delta$PBE0 shows significant systematic errors, particularly for the OH radical. $\Delta$PBE0 predicts an excitation energy of 535.1 eV for OH, a massive deviation of 9.3 eV from the experimental value of 525.8 eV. This single outlier heavily contributes to its high RMSE of 3.2 eV. While $\Delta$PBE0 performs adequately for some species, such as the CO$^+$ transitions where it is within 0.1–0.4 eV of the experiment, its instability in handling the OH radical suggests limitations in its self-interaction error correction or its description of relaxation effects upon core-hole creation for this specific open-shell system.

While USTEOM-CCSD is generally considered a high-level method, its performance here is mixed. It yields a MAE of 1.3 eV and an RMSE of 1.1 eV, placing it behind OBMP2 in terms of overall accuracy. Notably, for the OH radical, USTEOM-CCSD predicts an excitation energy of 524.5 eV against the experimental 525.8 eV, a reasonable agreement, but it fails to outperform the simpler UHF method in several instances. Surprisingly, the UHF method performs robustly with a MAE of 1.0 eV and RMSE of 1.1 eV. Although UHF tends to overestimate certain transitions, such as the prediction of 527.4 eV for OH compared to the experimental value of 525.8 eV, its error cancelation appears fortuitous in this specific test set, yielding results comparable to the more computationally expensive coupled-cluster method.

The OBMP2 method consistently provides predictions closest to the experimental benchmarks across the majority of the dataset. For instance, in the NO molecule, the OBMP2 value of 399.0 eV is very close to the experimental value of 399.7 eV, whereas USTEOM-CCSD underestimates this transition significantly at 397.4 eV. Similarly, for the CO$^+$ cation, OBMP2 predicts the 282.0 eV transition at 281.1 eV, a deviation of less than 1.0 eV, while UHF overestimates it at 283.5 eV. This trend of high-fidelity reproduction suggests that OBMP2 effectively balances correlation effects required for accurate core-hole description.

\begin{table*}[t!]
  \normalsize
  \caption{K-Edge Excitation Energies (eV) of Open-Shell Molecules Calculated from $\Delta$OBMP2 and $\Delta$PBE0. Results from CVS-USTEOM-CCSD are adapted from Ref~\cite{casanova2025core}. Experimental values are taken from Ref.~\cite{NO_O2_Kedge} for NO, Ref.~\cite{CO_ca_Kedge} for CO$^+$, Ref.~\cite{OH_Kedge} for OH, Ref.~\cite{NH_ca_Kedge} for NH$_2^+$ and NH. \label{tab:open-shell} \normalsize }
  \begin{tabular}{ccccccccccccccccccccc}
    \hline \hline		
    Molecules &Sym       &Exp$^a$ &UHF	&OBMP2	&PBE0  &USTEOM-CCSD\cite{casanova2025core} \\
    \hline
    \textbf{N}O		  &$2^2B_1$    &399.7	&399.2	&399.0	&398.2	&397.4 \\
	N\textbf{O}      &$2^2B_1$     &532.7	 &531.5	 &532.8	 &531.3	 &532.0 \\
	       &$3^2B_1$    &540.2	 &539.2	 &540.0	 &532.9	 &540.0 \\
    \textbf{C}O$^+$		  &$2^2A_1$    &282.0	&283.5	&281.1	&281.9	&280.5 \\
		       &$1^2(B_1 + B_2)$     &289.9	 &289.7	 &288.9	 &288.6	  &287.5 \\
	C\textbf{O}$^+$         &$2^2A_1$    &528.5   &528.8  &529.2  &528.1  &529.7 \\
		       &$1^2(B_1 + B_2)$     &533.4	 &535.2	 &536.0	 &534.2	  &534.1 \\
    \textbf{O}H		  &$A_1$      &525.8	&527.4	&526.0	&535.1	&524.5 \\
    \textbf{N}H$_2^+$  &$1^3B_1$	  &396.2	&397.2	&395.9	&395.6	 &-   \\
		       &$1^3A_1$   &396.4	 &397.5	 &396.2	 &395.9	  &-    \\
		       &$2^3B_1$   &407.2	 &407.2	 &406.8	 &406.1	  &-    \\
    \textbf{N}H$^+$	  &$2^4A_2$	  &394.9	&395.7	&394.7	&394.3	&394.7 \\
		     .  .&$1^4(B_1+B2)$ &397.8	 &398.8	 &397.8	  &397.2	&397.6 \\
		       &$1^A_2$    &398.8	 &397.4	 &397.0	  &396.2	  &395.0 \\
		       &$2^A_2$    &398.8	 &400.1	 &399.0	  &398.7	  &398.2 \\
\hline \hline
\end{tabular}
\end{table*}

\section{Conclusion}

In conclusion, we have evaluated the performance of the OBMP2-based $\Delta$SCF approach for the prediction of K-edge excited states. By employing test sets comprising both closed- and open-shell molecules, we demonstrated that this method consistently outperforms standard techniques, including $\Delta$DFT, EOM-CCSD, and USTEOM-CCSD. These findings highlight the efficacy of including self-consistent MP2 correlation in the orbital optimization process. Consequently, the OBMP2-based $\Delta$SCF protocol stands as a robust and accurate new computational strategy for the theoretical treatment of core-level spectroscopy.

\section*{Acknowledgement}
This research is funded by University of Science, VNU-HCM under grant number T2024-94.

\bibliography{main}

\end{document}